\pdfoutput=1

\documentclass[aps,prl,reprint,showpacs,superscriptaddress]{revtex4-1}
\usepackage{graphicx}
\usepackage{graphics}
\usepackage{amsmath}
\usepackage{amssymb}
\usepackage{amsfonts}
\usepackage{dcolumn}
\usepackage{dsfont}
\usepackage{latexsym}
\usepackage{rotating}
\usepackage{color}
\usepackage{latexsym}
\usepackage{bbm}
\usepackage{subfigure}
\usepackage{float}
\usepackage{epsfig}
\usepackage{epsf}
\usepackage{psfrag}
\usepackage{bm}
\usepackage{amsthm}
\usepackage{eucal}
\usepackage{mathrsfs}
\usepackage{url}
\usepackage{braket}
\usepackage{epstopdf}

\usepackage{times}
\usepackage{mathtools}
\usepackage{mathcomp}
\usepackage{soul}
\usepackage{tabularx}
\usepackage{calrsfs}
\usepackage[cp1252]{inputenc}
\usepackage[normalem]{ulem}
\usepackage[vcentermath]{youngtab}

\usepackage{adjustbox}

\usepackage{color} 

\newcommand{\dn}{\downarrow}

\newcommand{\up}{\uparrow}

\newcommand{\enni}{\noindent}
\newcommand{\enbe}{\begin{equation}}
\newcommand{\enee}{\end{equation}}
\newcommand{\enba}{\begin{align}}
\newcommand{\enea}{\end{align}}

\usepackage{hyperref}
\hypersetup{
colorlinks=true,final=true,
        linkcolor=blue,
        citecolor=blue,
        filecolor=blue,
        urlcolor=blue,
}
  
\begin{document}
  
\title{Theoretical investigation of superconductivity in trilayer square-planar nickelates}

\author{Emilian M. Nica}  
\email[Corresponding author: ]{enica@asu.edu}
\affiliation{Department of Physics, Arizona State University Tempe, Arizona 85287-1504, USA}

\author{Jyoti Krishna}
\affiliation{Department of Physics, Arizona State University Tempe, Arizona 85287-1504, USA}

\author{Rong Yu}
\affiliation{Department of Physics, Renmin University of China, 59 Zhongguancun St, Beijing, China, 100872}

\author{Qimiao Si}
\affiliation{Department of Physics and Astronomy, Rice University, 6100 Main St, Houston, TX, 77005, USA}
\affiliation{Rice Center for Quantum Materials, Rice University, 6100 Main St, Houston 77005 TX, USA}
\date{\today}

\author{Antia S. Botana}
\email[Corresponding author: ]{antia.botana@asu.edu}
\affiliation{Department of Physics, Arizona State University Tempe, Arizona 85287-1504, USA}

\author{Onur Erten}
\affiliation{Department of Physics, Arizona State University Tempe, Arizona 85287-1504, USA}


  
\begin{abstract}
The discovery of superconductivity in Sr-doped NdNiO$_{2}$ is a crucial breakthrough in the long pursuit for nickel oxide materials with electronic and magnetic properties similar to those of the cuprates. NdNiO$_2$ is the infinite-layer
 member  of a family of square-planar nickelates with  general chemical formula  R$_{n+1}$Ni$_n$O$_{2n+2}$ (R = La, Pr, Nd, $n= 2, 3, ... \infty$). In this letter, we investigate 
superconductivity in the trilayer member of this series (R$_4$Ni$_3$O$_8$) using a combination of first-principles and $t-J$ model calculations. 
 R$_4$Ni$_3$O$_8$ compounds resemble  cuprates more than RNiO$_2$ materials in that only Ni-$d_{x^{2}-y^{2}}$ bands cross the Fermi level, they exhibit a largely reduced charge transfer energy, and as a consequence  superexchange interactions are significantly enhanced. We find that the superconducting instability in doped R$_4$Ni$_3$O$_8$ compounds is considerably stronger with a maximum gap about four times larger than that in Sr$_{0.2}$Nd$_{0.8}$NiO$_2$.
 
\end{abstract}
 
\maketitle
\noindent
Understanding the mechanism behind high-temperature superconductivity (HTS) in the cuprate family remains one of the main challenges in condensed matter physics~\cite{norman-RPP}. One way of addressing this open question has been to search for cuprate analogs that display the electronic and magnetic properties deemed relevant to HTS: a layered structure similar to that of the CuO$_2$ planes, 
$d^{9}$ spin-1/2 ions, strong antiferromagnetic (AFM) correlations, isolated $d_{x^2-y^2}$ bands near the Fermi energy, and strong $p-d$ hybridization. In this regard, Ni-based compounds have been promising candidates since Ni$^{1+}$ and Cu$^{2+}$ are isoelectronic $d^{9}$ ions~\cite{anisimov}. A Ni$^{1+}$ oxidation state is indeed realized in square-planar infinite-layer systems RNiO$_2$ (R = La, Nd). After more than three decades of effort~\cite{crespin,hayward,hayward_nd,ikeda,ikeda2}, superconductivity in NdNiO$_2$ (112) was recently observed upon Sr-doping with $T_c \sim 15 $K~\cite{new}. 

First-principles calculations based on density-functional
theory (DFT) reveal similarities as well as differences between the parent infinite-layer 112 compound NdNiO$_{2}$
and members of the cuprate family \cite{pickett, prx, arita, deveraux, doped_MI, Thomale_PRB2020,eff_ham, ES_112, lechermann, pickett2}. While a single Ni-$d_{x^2-y^2}$ band indeed crosses the Fermi level as in the cuprates, 
R-$d$ electron pockets are also present. These likely prevent the parent phase from becoming a simple Mott insulator and suggest that Kondo interactions may play a relevant role ~\cite{deveraux}. Nonetheless, Sr-doped 112 compounds have a Fermi surface which is 
similar to that of the cuprates and several authors have proposed an analogous $d_{x^2-y^2}$ pairing order parameter~\cite{Thomale_PRB2020, Sakakibara_arxiv2019}.
 
\begin{figure}[t!]
	\includegraphics[scale=0.5]{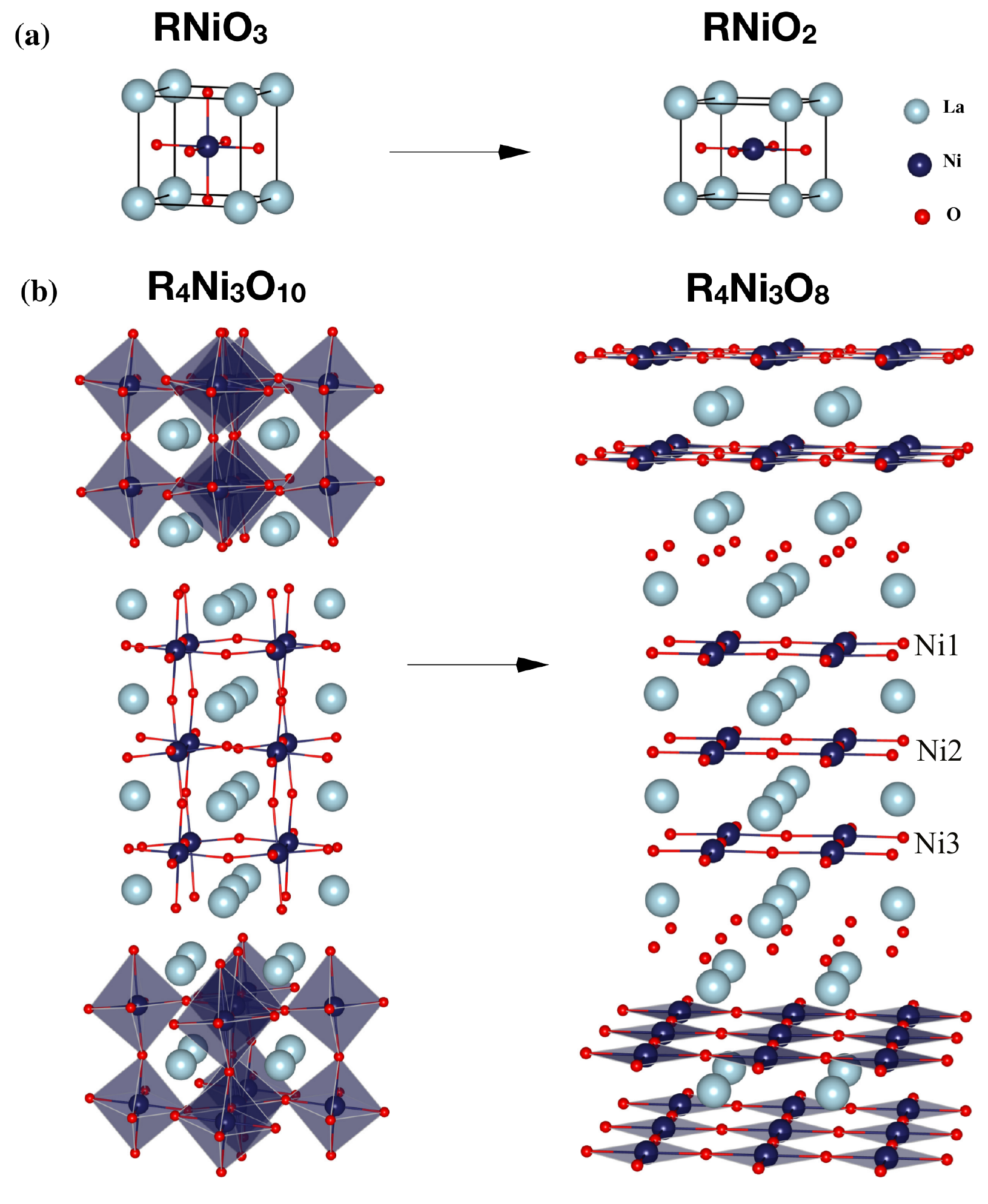}
	\caption{\label{fig1} Crystal structure of 112 and 438 square-planar nickelates illustrated in (a) and (b) panels on the right. These compounds
	are obtained via oxygen reduction from the corresponding 113 and 4310 perovskite-like parent compounds illustrated in the panels on the left.  Oxygen, rare-earth (R), and nickel atoms are depicted in red, light blue, and dark blue, respectively.}
	\label{Fig1}
\end{figure}

Importantly, 112 nickelates are the infinite-layer members of a larger series represented by the general formula
R$_{n+1}$Ni$_n$O$_{2n+2}$ (R = La, Pr, Nd, $n= 2; 3; ... \infty$) with each member containing $n$-NiO$_2$ layers~\cite{poltavets1, poltavets2}. The materials in this series are obtained via oxygen reduction from perovskite-like parent phases~\cite{poltavets1, poltavets2} as shown in Fig. 1. The fact that 112 nickelates belong to this larger series suggests the existence of a cuprate-like family of nickelate HTS.

Among the other members of this nickelate family, the trilayer materials R$_4$Ni$_3$O$_8$ (438) 
and especially Pr$_4$Ni$_3$O$_8$ have already been defined as close analogs of the cuprates, and therefore, promising candidates for HTS~\cite{nat_phys, physrevmat}. The structure of the $n=3$ and $n=\infty$ layered materials and their corresponding parent phases are shown in Fig. \ref{fig1}. 
In 112 compounds, adjacent NiO$_2$ planes are separated by a single layer formed by the rare-earth ions. 438 compounds exhibit trilayer blocks with an analogous structure. However, each of these blocks is separated along the $c$-axis by a fluorite slab formed by the rare-earth and oxygen ions. 
For the 438 materials, an average Ni valence of 1.33+ ($d^{8.67}$) is obtained~\cite{nat_phys}. In terms of $d$ filling, 438 compounds can be mapped onto the overdoped regime of the cuprate phase diagram ~\cite{Lee_2006} suggesting that HTS would likely be accessible via electron-doping. To determine if this is the case, we study the electronic structure and superconductivity in a $t-J$ model for the 438 trilayer nickelates. We also analyze a related $t-J$ model for the 112 compounds in order to provide a reference within the Ni-based family. We find robust $d_{x^2-y^2}$-wave superconductivity in the 438 materials upon electron-doping, with a maximum $T_{c} \sim 90 $K, much larger than that displayed by the 112 material as a consequence of an enhanced superexchange interaction. 

We performed density functional theory (DFT)-based calculations for 438 and 112 nickelates using the all-electron, full potential code WIEN2k~\cite{wien2k} based on the augmented plane wave plus local orbitals (APW + lo) basis set. The Perdew-Burke-Ernzerhof version of the generalized gradient approximation (GGA)~\cite{pbe} was used for the paramagnetic calculations. More details on the simulations are provided in Ref.~\onlinecite{suppl}. 
In order to avoid issues connected with Nd- and Pr-$4f$ states, we perform calculations for LaNiO$_2$ with lattice parameters adopted from NdNiO$_2$ and La$_4$Ni$_3$O$_8$, likewise with lattice parameters adopted from Pr$_4$Ni$_3$O$_8$. The extraction of tight binding (TB) parameters 
for effective $t-J$ models is based on the Wannier functions formalism~\cite{wannier90,wien2wannier}. All hopping coefficients and on-site energies obtained from the Wannier fits for the 438 compounds are shown in Table I of Ref.~\onlinecite{suppl}.

\begin{figure}[t]
\centering
\includegraphics[width=0.45\textwidth]{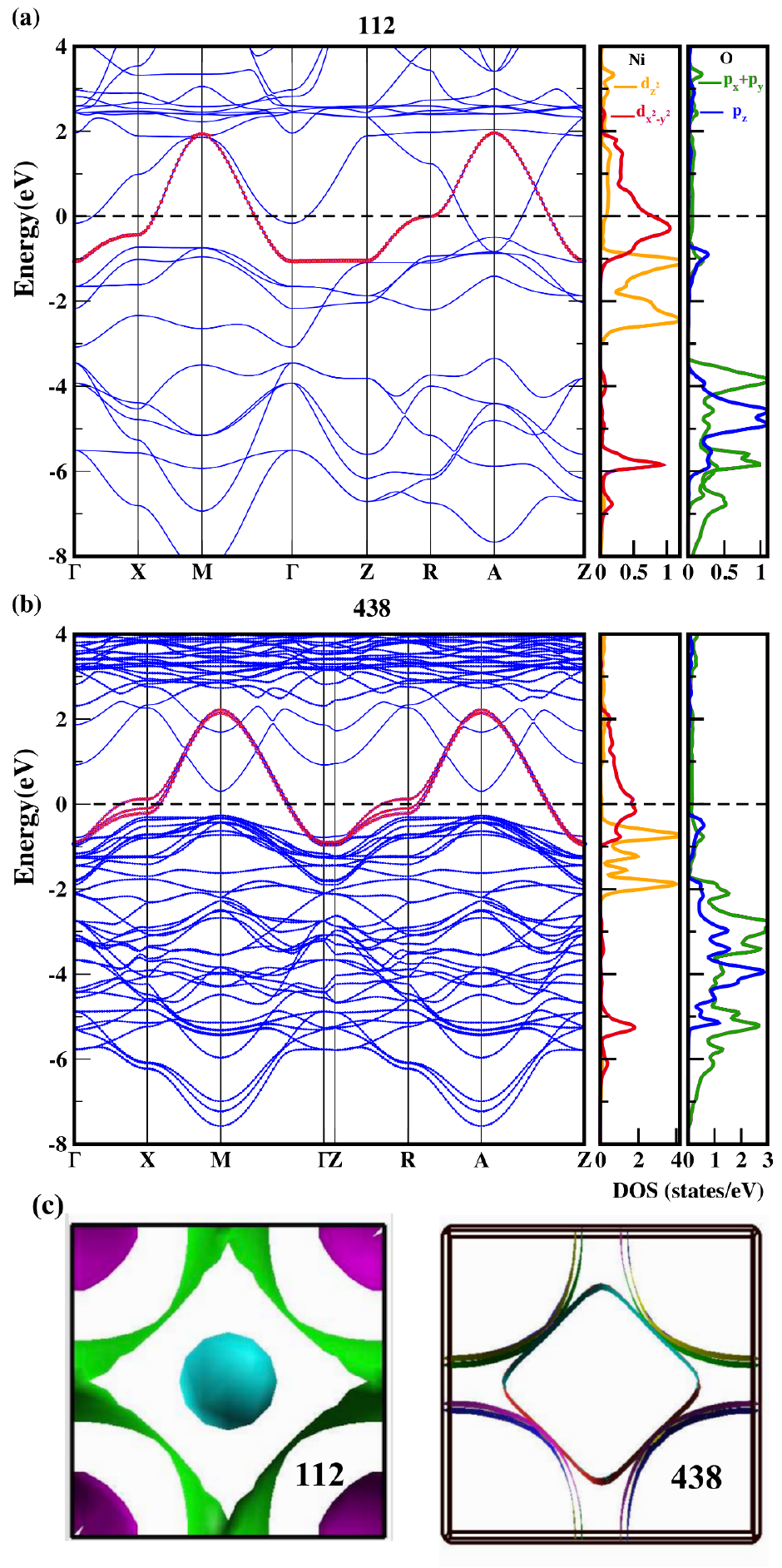}
 \caption{\label{fig2} Band structure and orbital projected DOS for (a) 112 and (b) 438 compounds. 
  Bands with dominant Ni $d_{x^{2}-y^{2}}$ character crossing the Fermi level are shown in red. The  right panels show the Ni $d_{x^{2}-y^{2}}$ and $d_{z^{2}}$ and O $p_{x}, p_{y}$, and $p_{z}$ orbital projected DOS. Panel (c) shows the corresponding Fermi surfaces.}
\end{figure}

Fig. \ref{fig2} shows the paramagnetic band structures and orbital-projected density of states (DOS) of La112 ($d^{9}$) and La438 ($d^{8.67}$) materials. In the 112 case shown in Fig.~\ref{fig1}~(a) the Ni-$d_{x^2-y^2}$ band crosses the Fermi level. However, as determined in previous work \cite{pickett, prx, arita, deveraux, doped_MI, ES_112},
additional Nd-$5d$ bands also contribute to the Fermi surface, giving rise to two electron pockets that self-dope the $d_{x^2-y^2}$ band (see Fig. \ref{fig2}(c)). The pocket at $\Gamma$ 
has predominant Nd-$d_{z^2}$ character, while the pocket at A is due mainly to the Nd $d_{xy}$ orbital. The large separation in energy between O-$p$  and Ni-$d$ bands is apparent from the DOS, shown in the panels to the right. The corresponding charge transfer energy $\Delta$= $E_{d}$- $E_{p}$ is derived from the values of the on-site energies for the Wannier functions as $\Delta_{112}$ $\sim$ 4.4 eV. 

The band structure for the 438 compounds shown in Fig.~\ref{fig2}~(b) differs significantly from that of the 112's.
Only a single Ni $d_{x^{2}-y^{2}}$ band per Ni crosses the Fermi level in analogy to the cuprates but in sharp contrast to the 112 compounds. 
The rare-earth $d$ bands are displaced to roughly 0.5 eV above the Fermi level. A splitting between the three Ni-$d_{x^{2}-y^{2}}$ bands is observed at $X$ as a consequence of interlayer hopping similar to that in multilayer cuprates~\cite{Lan2008}. The corresponding Fermi surface  (shown in Fig. \ref{fig2}(c)) resembles that of heavily hole-doped cuprates bearing one electron pocket (coming from the inner Ni) and two hole-pockets (from the outer ones).
More importantly, the charge transfer energy is largely reduced in the 438 compounds, $\Delta_{438}\sim 3.4 $ eV, 1 eV smaller than that of the 112 material. The difference in charge transfer energies between the 112 and 438 cases is consistent with X-ray absorption experiments~\cite{nat_phys}. While a pre-peak is observed in the 438 materials at the O-K edge, indicative of oxygen holes, no such pre-peak is observed in the 112 materials \cite{deveraux}. 
The in-plane $p-d$ hopping coefficients in 112 \cite{prx} and 438 materials (see Ref. \cite{suppl}) are almost identical.

The difference in charge transfer energies between the 438 and 112 materials significantly impacts the corresponding  nearest-neighbor (NN) 
superexchange coupling $J$.  For 112 materials, the $J$  determined by experiments \cite{xps} is $\sim$ 25 meV, a quarter of the value characteristic of the cuprates. Here, we estimate $J$ in the 438 case theoretically using an expression~\cite{khomskii_2014} which includes 
both the Mott and charge transfer limits: 
\begin{equation}
    J_{dd}= \frac{2t_{pd}^4}{\Delta^2U_{dd}}+ \frac{4t_{pd}^4}{\Delta^2(2\Delta+ U_{pp})}.
\end{equation}

Our estimates of $\Delta_{438}$ and t$_{pd}$ together with values of U$_{pp}$ and U$_{dd}$ characteristic of the cuprates \cite{martin}, lead to $J \approx$ 80 meV.  Ab initio calculations  show  that Ce-doped Pr$_4$Ni$_3$O$_8$ has the same electronic structure as the antiferromagnetic insulating phase of parent cuprates at half filling. For this system (CePr$_3$Ni$_3$O$_8$, $d^9$)  a similar value of $J$ is derived by fitting  the energies of different magnetic configurations to a Heisenberg model \cite{physrevmat}. 

Having established the relevant orbital content from the DFT bands and the superexchange couplings, we now consider
effective $t-J$ models for these two materials. 
While our focus is on the 438 case, we also provide results for the 112 case, as a reference within the Ni-based family.
For the 438 case, we consider an effective three Ni-$d$ orbital $t-J$ model:

\begin{eqnarray}
H&=&P_{s} \left( H_t+H_J \right) P_{s} \nonumber \\
H_{t} &=& -\sum_{\substack{ij \\ \alpha \beta, \sigma}}  \left(t^{\alpha \beta}_{ij} c^{\dag}_{i \alpha \sigma} c_{j \beta \sigma} + \text{h.c.} \right) - \mu \sum_{i,\alpha,\sigma}  c^{\dag}_{i \alpha \sigma} c_{i \alpha \sigma}  \label{Eq:3ly_Ht} \nonumber\\
H_{J} &= &  \sum_{\braket{ij}, \alpha} J^{\alpha \alpha}_{ij} \left( \bm{S}_{i \alpha} \cdot \bm{S}_{j \alpha} - \frac{n_{i\alpha} n_{j\alpha}}{4} \right) \nonumber\\
&+& \sum_{\alpha < \beta, i} J^{\alpha \beta}_{ii} \left( \bm{S}_{i \alpha} \cdot \bm{S}_{i \beta} - \frac{n_{i\alpha} n_{i\beta}}{4} \right).
\end{eqnarray}

\begin{figure}[t!]
\resizebox{0.95\columnwidth}{!}{\input{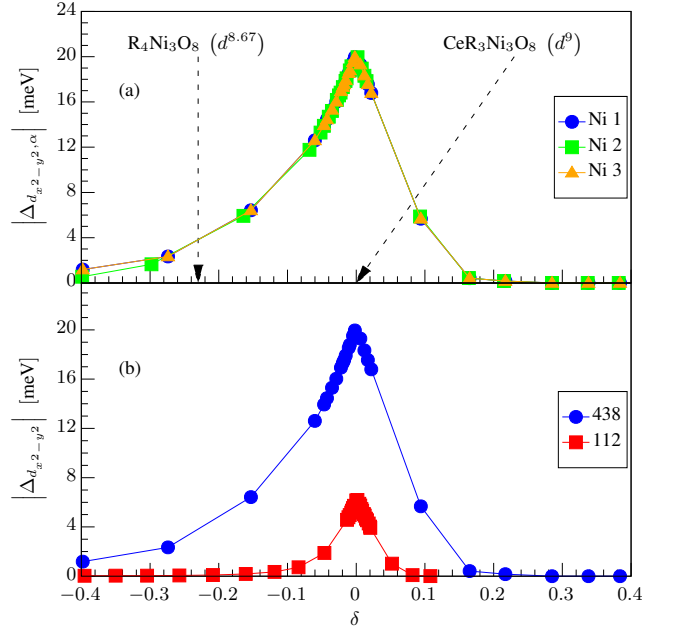}}
\caption{(a) Pairing amplitudes at $T=0$ for the 438 compounds in the $d_{x^{2}-y^{2}}$ channel as a function of doping for each of the Ni orbitals as determined from Eq.~\ref{Eq:Prng_ampl}. The arrows show the compound at half filling CeR$_{3}$Ni$_{3}$O$_{8}$ ($d^{9}$) and the parent phase R$_4$Ni$_{3}$O$_8$ ($d^{8.67}$). Ni1 and Ni3 occupy the outer layers while Ni2 occupies the inner layer as shown in Fig. \ref{fig1}. 
 The amplitudes in each Ni sector are similar. The offsets at lower hole-dopings are due to inter-layer coupling. ~(b) Pairing amplitude at $T=0$ for the 438 family in the $d_{x^{2}-y^{2}}$ channel in the Ni sector (blue circles) and pairing amplitude for the 112 family in the $d_{x^{2}-y^{2}}$ channel for Ni (red squares). Both are plotted as functions of their respective dopings. The leading pairing channel for the 438 case is roughly four times larger than that for the 112 case. }
\label{Fig:1}
\end{figure}
\begin{figure}[t!]
\resizebox{1.0\columnwidth}{!}{\input{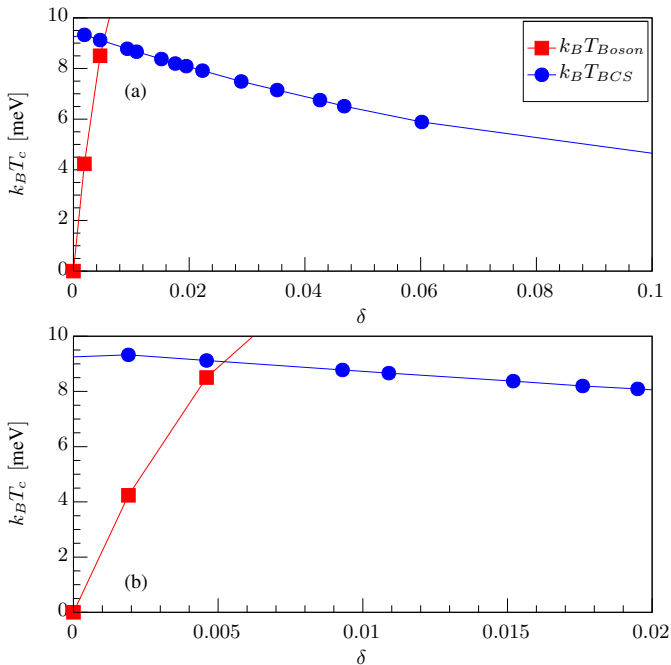}}
\caption{(a) Estimate of the superconducting transition temperature $T_{c}$ as a function of 
hole-doping from half filling ($d^9$)
 for the 438 compounds. The red squares give the estimate based on boson condensation temperature. The blue circles represent the estimate based on weak-coupling BCS theory. For a detailed account of the methods used to obtain the two estimates please consult Ref.~\onlinecite{suppl}.~ (b) Close-up view of (a). Note that the narrow under-doped region is due to the relatively large intra-layer NN  hopping as compared to the intra-layer exchange $J_{\parallel}$. A similar plot is obtained for 
 electron-doping.}
\label{Fig:2}
\end{figure} 
\enni The indices $i,j$ cover all of the sites of a two-dimensional square lattice, as the weak dispersion along the $c$-axis is neglected. The indices $\alpha, \beta \in\ \{1,2,3\}$ label the Ni $d_{x^{2}-y^{2}}$ orbitals in each of the three NiO$_{2}$ layers. The projection operator $P_{s}$ enforces the exclusion of doubly-occupied states in each Ni sector for hole-doping. A similar model is used for electron-doping, as discussed in Ref. ~\onlinecite{suppl}. 

In order to capture the DFT bands, NN, next-NN (NNN) and third-NN in-plane hopping terms are included $t^{\alpha \alpha}_{xx}=390~{\rm meV}, t^{\alpha \alpha}_{xy}=-100~{\rm meV}, t^{\alpha \alpha}_{xxy}=41~{\rm meV}$, which are identical for all Ni sectors. The splitting between the three Ni bands is due to an inter-layer hybridization $-2t^{\alpha \beta} \left( \cos(k_{x}a) - \cos(k_{y}a) \right)^{2}$, where $t^{\alpha \beta} = 16$ meV is used as obtained from DFT calculations for $(\alpha , \beta) \in \{(1,2), (2,3) \}$. 
These inter-layer terms are typical in multilayered cuprates~\cite{suppl}.
As NN intra-layer exchange we use the $J^{\alpha \alpha}_{ij}=J_{\parallel}=80$ meV value derived previously. 
Local, inter-layer exchanges are also included $J^{\alpha \beta}_{ii}=J_{z}=17$ meV for $(\alpha , \beta) \in \{(1,2), (2,3) \}$, identical for each pair of inter-layer Ni orbitals.

A slave-boson representation for each Ni  orbital is introduced ~\cite{Kotliar} and  solutions at $T=0$ which preserve time-reversal as well as all of the symmetries of the lattice are considered. In addition to the usual Gutzwiller condition, the filling in each Ni sector is also fixed, as determined from the non-interacting bands. This condition is the main approximation of our model. Consequently, at $T=0$, the boson in each sector $\braket{b^{\dag}_{i\alpha}}, \braket{b_{i\alpha}}$ can be replaced by $b_{\alpha} =\sqrt{\delta_{\alpha}}$, where $\delta_{\alpha}$ is the hole doping of the Ni $d_{x^{2}-y^{2}}$ in layer $\alpha$. Within a Hartree-Fock self-consistent approach, we also decouple the exchange interactions in NN intra-layer and local inter-layer particle-particle (p-p) and particle-hole (p-h) channels

\enni \begin{align}
B^{(\alpha \alpha)}_{\bm{e}} 
= & \frac{2}{N_{s}} \sum_{\bm{k}} \cos(\bm{k} \cdot \bm{e}) \braket{f_{\bm{k} \alpha \dn}f_{\bm{-k} \alpha \up}} \\ 
B^{(\alpha \beta)} = & \frac{1}{N_{s}} \sum_{\bm{k}} \braket{f_{\bm{k} \alpha \dn}f_{\bm{-k} \beta \up}+ f_{\bm{k} \beta \dn}f_{\bm{-k} \alpha \up}} \\
\chi^{(\alpha \alpha)}_{\bm{e}} 
= & \frac{1}{N_{s}} \sum_{\bm{k}} \sum_{\sigma} e^{i\bm{k} \cdot \bm{e}} \braket{f^{\dag}_{\bm{k} \alpha \sigma}f_{\bm{k} \alpha \sigma}} \\
\chi^{(\alpha \beta)} 
= &\frac{1}{N_{s}} \sum_{\bm{k}} \sum_{\sigma}  \braket{f^{\dag}_{\bm{k} \alpha \sigma}f_{\bm{k} \beta \sigma}},
\end{align}
\enni where $N_{s}$ denotes the number of sites of the 2D lattice, $\bm{e} \in \{ \bm{\hat{x}}, \bm{\hat{y}}\}$ is the NN intra-layer separation, and $f_{\bm{k} \alpha \sigma}$ is the electron operator in the $\alpha$ sector within the slave-boson representation. 

To study the effects of doping in the 112 material, we consider an effective $t-J$ model involving the Ni and two rare-earth Nd $d$ bands which cross the Fermi energy~\cite{Thomale_PRB2020}. Exchange couplings and projection of doubly-occupied states are in effect only for the Ni $d_{x^{2}-y^{2}}$ orbital. As in the 438 case, we fix the filling of the Ni orbital from the non-interacting bands, and decouple into p-p and p-h channels within a slave-boson representation. Our procedure incorporates the effects of interactions via band-renormalization~\cite{Kotliar} as in  similar approaches for the cuprates, while retaining realistic values of the exchange coupling constants. Discussions of this model and its solution are available in Ref.~\onlinecite{suppl}.

In Fig.~\ref{Fig:1}~(a), we present the pairing amplitude or gap order-parameters in the $d_{x^{2}-y^{2}}$ irreducible representation at $T=0$ for the three Ni orbitals in the 438 case as functions of their respective dopings $\delta_{\alpha}$  with respect to the half-filled system CeR$_{3}$Ni$_{3}$O$_{8}$ ($d^{9}$) 
. The position of the parent R$_4$Ni$_3$O$_8$ phase ($d^{8.67}$) is also shown. The gap order-parameters are determined from 

\begin{align}
\Delta_{d_{x^{2}-y^{2} \alpha}} = \frac{3J_{\parallel}}{4} \left( B^{\alpha \alpha}_{\bm{\hat{x}}} - B^{\alpha \alpha}_{\bm{\hat{y}}} \right). 
\label{Eq:Prng_ampl}
\end{align}

\enni The intra-layer pairings in the $s_{x^{2}+y^{2}}$ channels, as well as all of the inter-layer pairing channels are strongly suppressed in the doping regimes shown here. Similarly, all of the inter-layer $T=0$ p-h mean-field parameters  are suppressed relative to the intra-layer values (see Ref.~\onlinecite{suppl}). The amplitudes for $d_{x^{2}-y^{2}}$ pairing in the three Ni sectors shown in Fig.~\ref{Fig:1}~(a) follow a very similar evolution with doping. The slight anisotropy in hole- versus electron-doping can be traced to the p-h anisotropy of the bands shown in Fig.~\ref{fig2}~(b). Similarly, the distinction between the three Ni sectors for larger hole-dopings can be attributed to the distinct fillings in two of the orbitals versus the third. These different fillings are due to the inter-layer hopping and exchange coupling, together with the reflection symmetry about the middle plane. Within our approximations, small rare-earth pockets start to emerge beyond an electron doping of 0.1 relative to the 
CeR$_{3}$NiO$_{8}$ ($d^{9}$) configuration. As shown in Fig.~\ref{Fig:1}, the dominant pairing amplitude is already suppressed in this doping regime and we do not expect any  significant modifications to our results  due to these small pockets. 
We also note that the pairing in all three Ni sectors occurs with a zero relative phase, thus preserving the point-group and time-reversal symmetries.   In Fig.~\ref{Fig:1}~(b), the $d_{x^{2}-y^{2}}$ gap of the Ni sector is plotted for the 438's as a function of doping in comparison to that of the 112 materials. Remarkably,  the dominant pairing amplitude in the 438 systems is roughly four times larger than in the 112s.

In order to estimate the superconducting transition temperature $T_{c}$, we follow a procedure analogous to that for the $t-J$ model in the case of the cuprates~\cite{Kotliar, Lee_2006}. For the underdoped regime, $\delta_{\alpha} \rightarrow 0$, we estimate $T_{c}$ as the highest boson condensation temperature of the three Ni sectors. For the overdoped regime, $T_{c}$ is estimated via the value of the highest $d_{x^{2}-y^{2}}$ gap parameter at $T=0$ using weak-coupling Bardeen-Cooper-Schrieffer (BCS) theory~\cite{Won_Maki}. 

In Fig.~\ref{Fig:2}, we plot the estimates for the $T_{c}$ versus hole
-doping 
of 
the Ni orbitals with respect to the half-filled system CePr$_3$Ni$_3$O$_8$.
 The red line is the estimate based on boson condensation, while the blue line indicates the estimate based on weak-coupling BCS theory. The slope of the boson condensation temperature can also be estimated based on the analytical results for free bosons with weak dispersion along the $z$-axis, as in the case of the cuprates~\cite{Wen_Kan}. The very narrow under-doped region is due to a relatively large intra-layer NN hopping as compared to the intra-layer exchange coupling $J_{\parallel}$. For a more detailed account of these estimates please consult Ref~\onlinecite{suppl}. A similar plot is obtained for electron-doping. A maximum $T_{c} \sim 90 $~K upon hole-doping (from half filling) is found in the 438 materials, much larger than the $T_c\sim 15$~K observed in Sr-doped NdNiO$_2$. These results show that the $n=3$ (438) members of the layered nickelate family are even more promising candidates for superconductivity than the recently discovered infinite-layer superconductor (hole-doped 112). Both of them exhibit a dominant pairing instability in the d$_{x^2-y^2}$ channel but the $n=3$ material shows a pairing amplitude four times larger than that of the 112 systems and could potentially achieve a T$_c$ $\sim$ 90 K. 

To summarize, we have studied the electronic structure and superconducting instabilities of trilayer (438) nickelates and compared them to the recently-discovered infinite-layer superconductor (hole-doped 112). 
A DFT-based analysis of the 438 compounds indicates that these materials are much more cuprate-like than their 112 counterparts- they exhibit a superexchange interaction which approaches a value typical of the cuprate family (as a consequence of smaller $p-d$ charge transfer energy), and a single band of Ni-$d_{x^{2}-y^{2}}$ character is around the Fermi level (without rare-earth d bands). The solutions of the corresponding $t-J$ models at the mean-field level reveal that the dominant pairing in the $d_{x^{2}-y^{2}}$ channel is significantly stronger in the 438 versus 112 materials making these materials promising superconductors if electron doping can be achieved.

OE and ASB acknowledge NSF-DMR-1904716. EMN and JK acknowledge ASU for startup funds.  We acknowledge the ASU Research Computing Center for HPC resources. Work at Rice was supported by the DOE BES Award \# DE-SC0018197 and the Robert A. Welch Foundation Grant No. C-1411.

%

\pagebreak
\widetext
\begin{center}
\textbf{\large Supplementary Material for \\ ``Theoretical investigation of superconductivity in trilayer square-planar nickelates''
}
\end{center}

\makeatletter

In the supplemental material, we provide discussions of the DFT band structure and $t-J$ model calculations which support the results discussed in the main text.

\section{Band structure}

\subsection{Computational methods for DFT calculations}
Electronic structure calculations were performed using the all-electron, full potential code WIEN2k \cite{wien2k} based on the augmented plane wave plus local orbitals (APW + lo) basis set. The Perdew-Burke-Ernzerhof version of the generalized gradient approximation (GGA) \cite{pbe} was used for non-magnetic cases. 
To avoid issues connected with Nd and Pr- $4f$ states, we perform calculations for LaNiO$_2$ with the lattice parameters determined for NdNiO$_2$
and La$_4$Ni$_3$O$_8$ with the lattice parameters of Pr$_4$Ni$_3$O$_8$
Explicitly, for the 112 case,
these lattice parameters are 
$a$= 3.92 \AA, $c$= 3.28 \AA. 
For the 438 case, these are
$a$= 3.97 \AA, $c$= 26.10 \AA. 
For self-consistentency,
we used 90 and 1470 $\vec{k}$ points in the irreducible Brillouin zone of 112 and 438 cases
, respectively, with the plane wave cut-off parameter (R$_{mt}$K$_{max}$) set as 7.0. Muffin-tin 
radii of 2.5, 1.99, and 1.72 were used for La, Ni and O ions, respectively. 
To further understand the electronic structure we performed an analysis based on maximally localized Wannier functions
(MLWFs) \cite{mlwf}. For the spread
functional minimization, we used WANNIER90 \cite{wannier90}. Post-processing of MLWFs to generate tight-binding band structures, hopping integrals, and plots of Wannier orbitals was achieved with WIEN2WANNIER \cite{wien2wannier}. We obtain an excellent agreement between the band structure obtained from the Wannier function interpolation and that derived from the DFT calculations for the 438 materials. 

\subsection{Wannierization}

\begin{figure}[h]
\centering
\includegraphics[width = 0.3\textwidth, angle = 270]{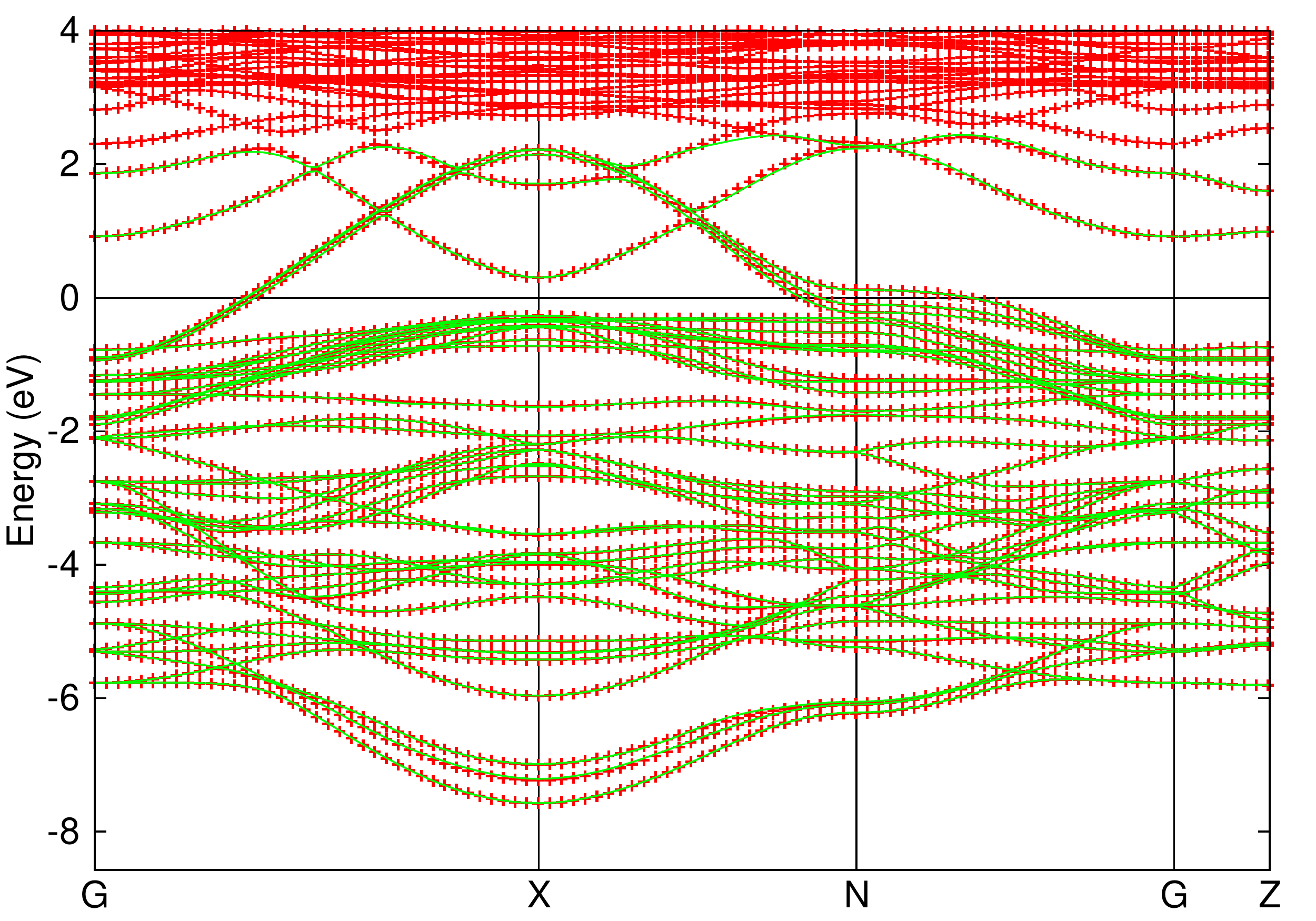}
 \caption{\label{Fig:S1} Comparison between the DFT (red) bandstructure and Wannier functions (green) interpolation with La $d_{z^{2}}$, La $d_{xy}$, Ni $d$ and O $p$ wavefunctions.}
\end{figure}

Excellent agreement is obtained between the band structure 
obtained from the Wannier function interpolation and that derived
from the DFT calculations for the 438 materials, showing
a faithful, albeit not unique, 
transformation to MLWFs. The Wannier functions describe $d$-like orbitals centered on the Ni sites which includes a small 
O-$p$ contribution for the $d_{x^2-y^2}$ orbitals
and $p$-like orbitals
on the O sites as shown in Fig.~\ref{Fig:S1}.
The spatial spread \cite{mlwf} of these functions is small and comparable in the Ni and Cu cases ($\sim$1 \AA$^2$). Table I gives the onsite energies and hopping integrals that are obtained from this process.

\begin{table}
\caption{Calculated on-site energies and hoppings for La$_4$Ni$_3$O$_8$ derived from the Wannier functions. Ni2 (inner) and Ni1 (outer), O1/O3 bonds to Ni inner/outer along the $x$ direction,
and O2/O4 bonds to Ni inner/outer along the $y$ direction. }
\begin{adjustbox}{width=0.8\columnwidth,center}
\begin{ruledtabular}
\begin{tabular}{lc}
\multicolumn{1}{l}{Wannier on-site energies} &
\multicolumn{1}{c}{ Value (eV)} \\
\hline

		d$_{xy}$ (Ni2) & -1.57\\
		d$_{yz/xz}$ (Ni2)&  -1.31\\
		 $d_{x^2-y^2}$ (Ni2) & -0.98\\
		d$_{z^{2}}$ (Ni2)& -1.16\\
		d$_{xy}$ (Ni2) & -1.53\\
		d$_{yz/xz}$ (Ni1/Ni3)&  -1.22\\
		$d_{x^2-y^2}$ (Ni1/Ni3) & -0.92\\
		d$_{z^{2}}$ (Ni1/Ni3)& -1.13 \\
		p$_{x}$(O1/O2) & -4.49 \\
		p$_{y}$(O1/O2) & -3.57 \\
		p$_{z}$ (O1/O2) & -3.50 \\
	   	p$_{x}$(O3/O4) & -4.47 \\
		p$_{y}$(O3/O4) & -3.57 \\
		p$_{z}$ (O3/O4) & -3.53 \\
		\hline
  Wannier hoppings (eV) &  \\ 
		\hline 
		p$_{y}$(O1/O2)-d$_{xy}$(Ni2) & -0.69\\
		p$_{z}$(O1)-d$_{yz}$ (Ni2)&  -0.76\\
		p$_{z}$(O2)-d$_{xz}$ (Ni2) & -0.72\\
		p$_{x}$(O1/O2)-$d_{x^2-y^2}$(Ni2) & -1.22 \\
		p$_{x}$(O1/O2)-d$_{z^{2}}$(Ni2)& -0.30 \\
		p$_{y}$(O3/O4)-d$_{xy}$(Ni1) & -0.68\\
		p$_{z}$(O3/O4)-d$_{yz}$ (Ni1)&  -0.72\\
		p$_{z}$(O3/O4)-d$_{xz}$ (Ni1) & -0.74\\
		p$_{x}$(O3/O4)-$d_{x^2-y^2}$(Ni1) & -1.22 \\
		p$_{x}$(O3/O4)-d$_{z^{2}}$(Ni1)& -0.28 \\
		p$_{z}$(O3/O4)-d$_{z^{2}}$(Ni1)& -0.12 \\
		p$_{y}$(O2)-p$_{x}$(O1)/ p$_{y}$(O4)-p$_{x}$(O3)& -0.26\\
		p$_{x}$(O1)-p$_{x}$(O2)/ p$_{x}$(O3)-p$_{x}$(O4)& -0.57\\
		p$_{y}$(O2)-p$_{y}$(O1)/ p$_{y}$(O3)-p$_{y}$(O4)& -0.20\\
			p$_{y}$(O1)-p$_{y}$(O3)/ p$_{y}$(O2)-p$_{y}$(O4) & -0.18\\
		p$_{z}$(O1)-p$_{y}$(O3) / p$_{z}$(O2)-p$_{y}$(O4) & -0.18\\	
	    p$_{z}$(O1)-p$_{z}$(O3) / p$_{z}$(O2)-p$_{z}$(O4) & -0.15\\	
	  	p$_{y}$(O1)-p$_{z}$(O3) / p$_{y}$(O2)-p$_{z}$(O4) & -0.15\\

\end{tabular}
\label{table2}
\end{ruledtabular}
\end{adjustbox}
\end{table}

\section{$t-J$ model and solution method for the 438 compounds}

\subsection{Mean-field solution at $T=0$ in the slave-boson formulation}

We use a slave-boson representation in order to exclude doubly-occupied states for each of the three Ni $d_{x^{2}-y^{2}}$ orbitals~\cite{Kotliar}. More specifically, we introduce one independent boson per each Ni sector together with the constraints

\enni \begin{align}
\sum_{\sigma}f^{\dag}_{i \alpha \sigma}f_{i \alpha \sigma} + b^{\dag}_{i \alpha} b_{i \alpha} = & 1 \label{Eq:3ly_Cndt_1}\\
\sum_{\sigma}f^{\dag}_{i \alpha \sigma}f_{i \alpha \sigma} = & 1-\delta_{\alpha}, \label{Eq:3ly_Cndt_2}
\end{align}

\enni where $\delta_{\alpha}$ are the dopings of the three Ni orbitals. To enforce these constraints we introduce the associated Lagrange multipliers $\lambda_{\alpha}$ and $\mu_{\alpha}$, respectively. 

While the first three constraints in Eq.~\ref{Eq:3ly_Cndt_1} are inherent to the slave-boson approach, the last three in Eq.~\ref{Eq:3ly_Cndt_2} fix each of the Ni fillings to given $\delta_{\alpha}$. Each $\delta_{\alpha}$ is determined from the  corresponding filling of the Ni $\alpha$ orbital at $T=0$: 

\enni \begin{align}
\delta_{\alpha} = &  1- n_{\alpha}.
\end{align}

\enni The orbital fillings are determined from the non-interacting bands as a function of chemical potential. 
In this approach, doping is equivalent to a rigid shift in chemical potential. We further discuss this approximation below. The Ni fillings are determined from

\enni \begin{align}
n_{\alpha} = &  \frac{1}{N_{s}} \sum_{a} \sum_{\bm{k}} \left|U_{\alpha a}(\bm{k}) \right|^{2} n_{F}\left( \epsilon_{\bm{k} a}  \right),
\end{align}

\enni where $N_{s}$ is the number of sites in the two-dimensional lattice, $a$ is a band index, the unitary matrix $U_{\alpha a}(\bm{k})$ diagonalizes the tight-binding (TB) Hamiltonian $H_{t}$ in Eq.~2 of the main text, $\epsilon_{\bm{k}a}$ are the band dispersions, and 

\enni \begin{align}
n_{F}\left( \epsilon_{\bm{k} a} \right) = \frac{1}{e^{\beta \left( \epsilon_{\bm{k} a} - \mu \right)}+1}
\end{align}

\enni is the Fermi-Dirac factor. The non-interacting TB Hamiltonian can be cast into the form 

\enni \begin{align}
H_{t} = & \sum_{\bm{k}} \sum_{\alpha \beta} \left[ \xi_{0}(\bm{k}) \lambda_{0} + \xi_{1}(\bm{k}) \left( \lambda_{1}+\lambda_{6} \right) \right]_{\alpha \beta} c^{\dag}_{\bm{k} \alpha} c_{\bm{k} \beta},
\label{Eq:Hmlt_bnds}
\end{align}

\enni where $\lambda_{0}$ is the $ 3 \times 3$ identity matrix while $\lambda_{1,6}$ are Gell-Mann matrices. It is straightforward to verify that $\left|U_{\alpha a}(\bm{k}) \right|$ will be identical for two of the orbitals. Consequently, two of the orbitals will have identical fillings while the third will generally be different for given chemical potential. 
The Ni fillings as a function of chemical potential are shown in Table~\ref{Table:2}.

\begin{table*}[h!]
\caption{Fillings of the three Ni orbitals $n_{\alpha}$ as functions of the chemical potential.}
\begin{ruledtabular}
\begin{tabular}{c c c c}
$\mu$ & $n_{1}$ & $n_{2}$ & $n_{3}$ \\
\hline 
-0.76 & 0.028 & 0.028 & 0.028 \\
-0.68 & 0.068 & 0.068 & 0.068 \\
-0.60 & 0.108 & 0.108 & 0.108 \\
-0.52 & 0.151 & 0.151 & 0.151 \\
-0.44 & 0.197 & 0.197 & 0.197 \\
-0.36 & 0.249 & 0.251 & 0.249 \\
-0.28 & 0.306 & 0.308 & 0.306 \\
-0.20 & 0.397 & 0.426 & 0.397 \\
-0.12 & 0.489 & 0.525 & 0.489 \\ 
-0.04 & 0.610 & 0.608 & 0.610 \\
0.04 & 0.718 & 0.695 & 0.718 \\
0.12 & 0.815 & 0.795 & 0.815 \\
0.20 & 0.903 & 0.897 & 0.903 \\
0.28 & 0.979 & 0.973 & 0.979 \\
0.36 & 1.040 & 1.035 & 1.040 \\
0.44 & 1.098 & 1.096 & 1.098 \\
0.52 & 1.157 & 1.155 & 1.157
\end{tabular}
\end{ruledtabular}
\label{Table:2}
\end{table*}

The $t-J$ model in the slave-boson representation is 

\enni \begin{align}
H_{438} = &  - \sum_{i \le j} \sum_{\alpha \le\beta} \sum_{\sigma}  t^{\alpha \beta}_{ij}
\left(  b_{i\alpha} b^{\dag}_{j \beta} f^{\dag}_{i \alpha \sigma} f_{j \beta \sigma} + h.c. \right)  
 \notag \\
-  & \frac{J_{\parallel} }{4} \sum_{\braket{ij}}\sum_{\alpha} \bigg[ \left( f^{\dag}_{i \alpha \up}f^{\dag}_{j \alpha \dn} - f^{\dag}_{i \alpha \dn}f^{\dag}_{j \alpha \up} \right)
\left( f_{i \alpha \dn}f_{j \alpha \up} - f_{i \alpha \up}f_{j \alpha \dn} \right) 
+ \sum_{\sigma \sigma'}  f^{\dag}_{i \alpha \sigma}  f_{j \alpha \sigma}  f^{\dag}_{j \alpha \sigma'} f_{i \alpha \sigma'} 
\notag \\
 - &  \sum_{\sigma}  f^{\dag}_{i \alpha \sigma} f_{i \alpha \sigma} 
+  \left(1-b^{\dag}_{i \alpha} b_{i \alpha} - b^{\dag}_{j \alpha} b_{j \alpha} - b^{\dag}_{i \alpha } b_{i \alpha} b^{\dag}_{j \alpha} b_{j \alpha} \right) \bigg]
\notag \\
-  & \frac{J_{z} }{4} \sum_{i}\sum_{\braket{\alpha \beta}} \bigg[ \left( f^{\dag}_{i \alpha \up}f^{\dag}_{i \beta \dn} - f^{\dag}_{i \alpha \dn}f^{\dag}_{i \beta \up} \right)
\left( f_{i \alpha \dn}f_{i \beta \up} - f_{i \alpha \up}f_{i \beta \dn} \right) 
+ \sum_{\sigma \sigma'}  f^{\dag}_{i \alpha \sigma}  f_{i \beta \sigma}  f^{\dag}_{i \beta \sigma'} f_{i \alpha \sigma'} 
\notag \\
 - &  \sum_{\sigma}  f^{\dag}_{i \alpha \sigma} f_{i \alpha \sigma} 
+  \left(1-b^{\dag}_{i \alpha} b_{i \alpha} - b^{\dag}_{i \beta} b_{i \beta} - b^{\dag}_{i \alpha } b_{i \alpha} b^{\dag}_{i \beta} b_{j \beta} \right) \bigg] 
\notag \\
+  & \sum_{i} \sum_{\alpha} \bigg[ \lambda_{i \alpha} \left( b^{\dag}_{i \alpha}b_{i \alpha} + \sum_{\sigma} f^{\dag}_{i \alpha  \sigma} f_{i \alpha \sigma} -1\right) 
-   \mu_{\alpha} \left( \sum_{\sigma} f^{\dag}_{i \alpha \sigma} f_{i \alpha \sigma} -1 + \delta_{\alpha} \right) \bigg].
\end{align} 

We consider uniform mean-field solutions at $T=0$ with $\braket{b^{\dag}_{i \alpha}}= b^{*}_{\alpha}, \braket{b_{i \alpha}}= b_{\alpha}$ and ignore terms which are bi-quadratic in the bosons. In contrast to a canonical single-band $t-J$ model~\cite{Kotliar}, $H_{438}$ also includes inter-layer hybridization terms $t^{\alpha \beta}_{ii}$, where $\alpha<\beta$. In general, the boson amplitudes and phases in each sector can be determined self-consistently for fixed \emph{total} filling by minimizing the appropriate Landau-Ginzburg (LG) action with respect to $b_{\alpha}$ and $b^{*}_{\alpha}$. The form of these self-consistency equations is similar to that of $H_{438-t}$ which determines the bands (Eq.~\ref{Eq:Hmlt_bnds}). One solution consists of real $b_{\alpha} \ge 0$ which  are identical for two of the orbitals. Our approximation, where the filling for each Ni orbital is fixed from the non-interacting bands, amounts to selecting the  self-consistent solution with all $b_{\alpha} \ge 0$. 

We decouple all of the exchange interactions in both particle-particle (p-p) and particle-hole (p-h) channels and absorb the $\lambda_{\alpha}$'s into the corresponding $\mu_{\alpha}$'s. We minimize the LG free energy per unit cell  

\enni \begin{align}
f = & -\frac{2T}{N_{s}} \sum_{m} \sum_{\bm{k}} \ln\left( \cosh(\beta E_{\bm{k}m})  \right) + \frac{3J_{\parallel}}{8} \sum_{\alpha} \left( \left|B^{(\alpha \alpha)}_{x} \right|^{2} + \left|B^{(\alpha \alpha)}_{y } \right|^{2} \right) + \frac{3J_{z}}{8} \left( \left|B^{(12)}\right|^{2} + \left|B^{(23)} \right|^{2} \right) 
\notag \\
+ & \sum_{\alpha} \frac{3J_{\parallel}}{8}\left(\left|\chi^{(\alpha \alpha)}_{x} \right|^{2} + \left| \chi^{(\alpha \alpha)}_{y} \right|^{2} \right) + \frac{3J_{z}}{8} \left( \left|\chi^{(12)}\right|^{2} + \left|\chi^{(23)} \right|^{2} \right) 
- \sum_{\alpha} \mu_{\alpha} \delta_{\alpha},
\end{align}  

\enni w.r.t. the intra-layer pairings $B^{(\alpha \alpha)}_{x,y,z}$, local inter-layer pairings $B^{(12)}, B^{(23)}$, intra-layer Hartree terms $K^{\alpha \alpha}_{x,y,z}$ and inter-layer Hartree terms $K^{(12)}, K^{(23)}$ at fixed filling for each of the three Ni orbitals. The uniform mean-field pairing terms are defined as

\enni \begin{align}
B^{(\alpha \alpha)}_{\bm{e} \alpha} = &  \braket{  f_{\bm{r}_{i}+\bm{e} \alpha \dn} f_{\bm{r}_{i} \alpha \up} - f_{\bm{r}_{i} + \bm{e}_{j} \alpha \up} f_{\bm{r}_{i} \alpha \dn}}  \notag \\
= & \frac{2}{N_{s}} \sum_{\bm{k}} \cos(\bm{k} \cdot \bm{e}) \braket{f_{\bm{k} \alpha \dn}f_{\bm{-k} \alpha \up}}, 
\label{Eq:Trly_B}
\end{align}

\enni where $\bm{e} \in \{ \bm{\hat{x}}, \bm{\hat{y}}\}$ and 

\enni \begin{align}
B^{(\alpha \beta)} = & \braket{f_{\bm{r} \alpha \dn}f_{\bm{r}_{i} \beta \up} - f_{\bm{r}_{i} \beta \up}f_{\bm{r}_{i} \alpha \dn} } \notag \\
= & \frac{1}{N_{s}} \sum_{\bm{k}} \braket{f_{\bm{k} \alpha \dn}f_{\bm{-k} \beta \up}+ f_{\bm{k} \beta \dn}f_{\bm{-k} \alpha \up}}.
\end{align}

\enni The uniform Hartree terms are 

\enni \begin{align}
\chi^{(\alpha \alpha)}_{\bm{e}} = & \sum_{\sigma} \braket{f^{\dag}_{\bm{r}_{i} \alpha \sigma}f_{\bm{r}_{i} + \bm{e} \alpha \sigma}}  
\notag \\
= & \frac{1}{N_{s}} \sum_{\bm{k}} \sum_{\sigma} e^{i\bm{k} \cdot \bm{e}} \braket{f^{\dag}_{\bm{k} \alpha \sigma}f_{\bm{k} \alpha \sigma}}, \label{Eq:Trly_K_1}
\end{align}

\enni and 

\enni \begin{align}
\chi^{(\alpha \beta)} = & \sum_{\sigma} \braket{f^{\dag}_{\bm{r}_{i} \alpha \sigma}f_{\bm{r}_{i} \beta \sigma}} \notag \\
= &\frac{1}{N_{s}} \sum_{\bm{k}} \sum_{\sigma}  \braket{f^{\dag}_{\bm{k} \alpha \sigma}f_{\bm{k} \beta \sigma}}
\label{Eq:Trly_K_2}
\end{align}

\enni $E_{\bm{k}m}$ are eigenvalues of 

\enni \begin{align}
H_{\bm{k}} = \begin{pmatrix}
h_{\bm{k};\alpha \beta}  & \Delta_{\bm{k}; \alpha \beta} \\
\Delta^{\dag}_{\bm{k}; \alpha \beta}  & - h^{T}_{-\bm{k};\alpha \beta} 
\end{pmatrix} 
\end{align}

\enni in a Nambu basis with spinor 
$\Psi^{T}=(f_{\bm{k} \alpha \up}, f^{\dag}_{-\bm{k} \alpha \dn})$. The normal part is given by 

\enni \begin{align}
h_{\bm{k}; \alpha \beta} = \text{F.T.}  \left[ H_{438-t; \alpha \beta} (b_{\alpha}) \right] - \frac{3J_{\parallel}}{4} \sum_{\bm{e}} \sum_{\alpha} \sum_{\sigma}  \text{Re} \left( \chi^{(\alpha \alpha)}_{\bm{e}} e^{i \bm{k} \cdot \bm{e}} \right) \delta_{\alpha, \beta} 
- \frac{3J_{z}}{4} \left( \chi^{(\alpha \beta)} \delta_{\alpha, \beta-1} + \chi^{(\alpha \beta), *} \delta_{\alpha-1, \beta} \right)
- \mu_{\alpha} \delta_{\alpha, \beta},
\end{align}

\enni where the Kronecker $\delta_{\alpha, \beta}$ is not to be confused with the dopings $\delta_{\alpha}$. $H_{438-t; \alpha \beta} (b_{\alpha})$ is determined from $H_{438-t}$ with tight-binding coefficients re-scaled by the appropriate $\sqrt{b_{\alpha}}$ factors. The pairing part of $H_{\bm{k}}$ is determined by 

\enni \begin{align}
\Delta_{\bm{k} \alpha \beta} = - \sum_{\bm{e} \in \{\bm{\hat{x}}, \bm{\hat{y}} \}} \frac{3J_{\parallel}}{4} \cos(\bm{k} \cdot \bm{e}) B^{(\alpha \alpha)}_{\bm{e}} \delta_{\alpha \beta} 
- \frac{3J_{z}}{4} B^{(\alpha \beta)}  \left( \delta_{\alpha, \beta-1} + \delta_{\alpha-1, \beta} \right)   
\end{align}

For electron-doped cases, we apply the well-known p-h transformation~\cite{Gooding} 

\enni \begin{align}
P_{s} c_{i \alpha \sigma} P_{s}\rightarrow \tilde{P}_{s} \tilde{c}^{\dag}_{i \alpha \sigma} \tilde{P}_{s} \label{Eq:p_h}. 
\end{align} 

\enni for each of the three Ni $d_{x^{2}-y^{2}}$ orbitals. $\tilde{P}_{s}$ projects all doubly-occupied hole states in each Ni sector. This transformation implicitly assumes that the fillings of each Ni orbital are either below or above half-filling. This turns our to be the case for all of the cases considered here and we do not explicitly treat cases where one of more of the orbital is hole-doped while the remaining orbitals are electron-doped. The transformed $t-J$ model is obtained from the hole-doped Hamiltonian by changing the signs of all TB coefficients. We solve these electron-doped cases together with the constraints

\enni \begin{align}
\sum_{\sigma}\tilde{f}^{\dag}_{i \alpha \sigma} \tilde{f}_{i \alpha \sigma} = & 1-\tilde{\delta}_{\alpha}.
\end{align}

\enni 
where $\tilde{\delta}_{\alpha}$ are the electron dopings for each Ni orbital.

In either hole- or electron-doped cases, the self-consistent solution is obtained numerically on an $1000 \times 1000$ grid in the first 2D Brillouin Zone.  

\subsection{Estimate of $T_{c}$}

As in the single-band $t-J$ model~\cite{Kotliar}, we estimate the superconducting $T_{c}$ via the boson condensation temperature in the underdoped regime and a weak-coupling Bardeen-Cooper-Schrieffer (BCS) theory in the overdoped regime. 

In the underdoped case, we generalize the procedure in the single-orbital $t-J$ model~\cite{Kotliar} by allowing fluctuations in each of the three boson sectors and by decoupling the boson density-density interactions via a Hartree-Fock approximation. The effective boson Hamiltonian is given by

\enni \begin{align}
H_{438-B} = & - \sum_{\bm{k}} \sum_{\alpha} \left[  \sum_{\bm{e}}  2t_{B \bm{e} \alpha} \cos(\bm{k} \cdot \bm{e}) 
- \left( J_{\parallel} + J_{z} + \lambda_{\alpha} \right)  \right] b^{\dag}_{\bm{k} \alpha} b_{\bm{k} \alpha} 
- \frac{J_{z}}{4} \sum_{\bm{k}}\sum_{\alpha < \beta} \left( b^{\dag}_{\bm{k} \alpha} b_{\bm{k} \beta} + \text{h.c.} \right) \label{Eq:Bsn_Hmlt}
\end{align}

\enni together with the constraints 

\enni \begin{align}
\braket{b^{\dag}_{i\alpha} b_{j\beta}} = \sqrt{\delta_{\alpha} \delta_{\beta}}.
\end{align}

\enni The tight-binding coefficients are 

\enni \begin{align}
t_{B \bm{e} \alpha } =  t^{\alpha  \alpha}_{\bm{e}} \chi^{(\alpha \alpha)}_{\bm{e}} + \frac{J_{\parallel}}{4},
\end{align}.

\enni with $\bm{e} \in \{\bm{\hat{x}}, \bm{\hat{y}} \}$. Note that we have ignored the inter-layer hybridization which is proportional to $t_{\parallel} \chi^{(\alpha \beta)} \approx \text{O}(10^{-4})$ near half-filling as illustrated in Fig.~\ref{Fig:1_SM}.

The last term in Eq.~\ref{Eq:Bsn_Hmlt} is due to the inter-layer density-density interactions. The main effect within a Hartree-Fock decomposition is a $\bm{k}$-independent splitting of the three boson bands. This splitting can be absorbed into renormalized chemical potentials $\lambda_{\alpha}$. 

In order to obtain finite boson condensation temperatures, we include a nominal dispersion along z as

\enni \begin{align}
H_{438-t} \rightarrow H_{438-t} - \sum_{\bm{k}} \sum_{\alpha} 2t_{Bz} \cos(k_{z} a_{z}) b^{\dag}_{\bm{k} \alpha} b_{\bm{k} \alpha}
\end{align}

\enni where $t_{Bz} / t_{B\bm{e}\alpha} \approx 10^{-2}$ and $a_{z}$ is a NN separation along z.
 
The highest condensation temperature then occurs in the boson sector with highest $\delta_{\alpha}$. For simplicity, we take this to correspond to $\alpha=1$. Near the condensation point the chemical potential $\lambda_{1}$ vanishes as
 
\enni \begin{align}
\lambda \rightarrow 2 \sum_{\bm{e}} t_{B \bm{e}} - J_{\parallel}-J_{z}.
\end{align}

\enni $T_{c}=1/\beta_{c}$ is obtained from the boson density constraint

\enni \begin{align}
\delta_{1} = & \frac{1}{N_{s}} \sum_{\bm{k}} n_{B}(\epsilon_{k}),
\end{align}

\enni where  

\enni \begin{align}
n_{B \bm{k}} = & \frac{1}{e^{\beta_{c} \epsilon_{B \bm{k}}}-1},
\end{align}

\enni is the Bose-Einstein factor and $\epsilon_{\bm{k}}$ are the eigenvalues of $H_{438-B}$ with splitting ignored.

Deep in the underdoped regime, the evolution of $T_{c}$ with doping
 can also be estimated from the analytical results for free bosons with weak dispersion along z~\cite{Wen_Kan}:

\enni \begin{align}
k_{B} T_{c} \approx \frac{2\pi n a_{z}}{m \ln \left( k_{B}T_{c} M \nu a^{2}_{z} \right)},
\end{align}

\enni where $n$ is the boson density, $a_{z}$ is the NN separation along the z direction, $m$ is the intra-layer effective mass, $M$ is the effective mass along z, and $\nu \approx O(1)$ in the regime of interest~\cite{Wen_Kan}. Near half-filling, $T_{c} \rightarrow 0$, and we ignore the logarithmic corrections. $m$ is estimated from the bottom of the quasi-2D boson bands determined from $H_{438-B}$ as

\enni \begin{align}
m^{-1} \approx 2t_{B} a^{2},
\end{align} 

\enni where $a$ is the NN intra-layer separation. Using the NN intra-layer $t^{11}_{ij}=400$ meV, $\chi_{\bm{\hat{x}}}=\chi_{\bm{\hat{y}}} \approx 0.35$ at half-filling, we estimate 

\enni \begin{align}
k_{B} T_{c} \approx \left(1.76~\text{eV} \right) \delta.
\end{align}

\enni The prefactor is similar to the value of 1.95 eV extracted from the numerical results in Fig.~4 of the main text.

In the over-doped regime, we estimate $T_{c}$ from weak-coupling $d$-wave BCS theory~\cite{Won_Maki} as

\enni \begin{align}
T_{c} \approx \max\left( \left|\max \Delta_{d} \right| \right) /2.14
\end{align} 

\enni by using the maximum of the three gaps. 

\subsection{Hartree mean-field parameters at $T=0$}

\begin{figure}[t!]
\resizebox{0.5\columnwidth}{!}{\input{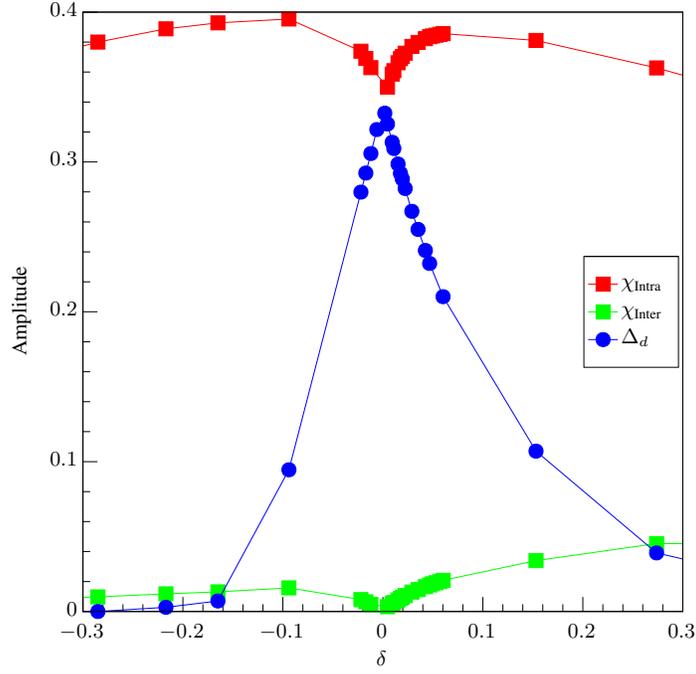}}
\caption{Dimensionless intra- and inter-layer Hartree mean-field parameters and dimensionless $d_{x^{2}-y^{2}}$ pairing amplitude for orbital 1 at $T=0$ as functions of doping for the same orbital for the 438 compounds.}
\label{Fig:1_SM}
\end{figure}

At $T=0$ the dimensionless Hartree terms are all real and obey

\enni \begin{align}
\chi^{(11)}_{\bm{\hat{x}}} = &\chi^{(11)}_{\bm{\hat{y}}} = \chi_{\text{Intra}} 
\\
\chi^{(12)} = & \chi^{(23)} = \chi_{\text{Inter}},
\end{align}

\enni where $\chi^{(\alpha \alpha)}_{\bm{\hat{e}}}$ and $\chi^{(\alpha \beta)}$ are defined in Eqs.~\ref{Eq:Trly_K_1} and~\ref{Eq:Trly_K_2}, respectively. The intra-plane components for the remaining 2,3 sectors behave similarly. 

We plot these dimensionless Hartree mean-field parameters alongside the dimensionless $d_{x^{2}-y^{2}}$ amplitude defined as

\enni \begin{align}
\Delta_{d} = & \left|B^{(11)}_{x} -B^{(11)}_{y} \right|,
\end{align} 

\enni where $B^{(\alpha \alpha)}_{\bm{\hat{e}}}$ is defined in Eq.~\ref{Eq:Trly_B}, as functions of the doping of orbital 1 in Fig.~\ref{Fig:1_SM}. We note that as we approach half-filling, $\chi_{\text{Intra}} = \Delta_{d}$ as in the case for the single-band $t-J$ model~\cite{Kotliar,Lee_2006}. The inter-layer Hartree terms are strongly suppressed near this point.

\section{$t-J$ model and solution method for the doped 112 compounds}
\subsection{Model}

The mechanism behind superconductivity in the 112 compounds is a matter of active debate. Nonetheless, Sr-doped NdNiO$_{2}$ provides the only known realization of Ni-based superconductivity. The parent compound in the 112 family is expected to have a Ni $d^{9}$ configuration based on an ionic count, which \emph{a priori} suggests the use of a $t-J$ model approach. Moreover, in order to compare the predicted superconducting instabilities in the 438 with those in the 112 compounds, we consider an effective three-orbital $t-J$ model for the latter which includes the Nd $5d$ $d_{xy}$ and $d_{z^{2}}$ orbitals in addition to the Ni 3d $d_{x^{2}-y^{2}}$. While strong correlations are in general important for all three orbitals based on their $d$-orbital nature, the Nd orbitals are expected to remain significantly away from half-filling throughout the doping range considered here. Therefore, we consider nearest-neighbor (NN) exchange interactions and impose the double-occupancy constraint exclusively on the Ni orbital. Likewise, we expect that the strongest pairing instability occurs in the Ni $d_{x^{2}-y^{2}}$ sector. Our effective $t-J$ model for the 112 compounds is

\enni \begin{align}
H_{112} = & H_{112-t} + H_{112-J} \label{Eq:Inly_H} \\
H_{112-t} = & P_{s} \left[ -\sum_{i<j} \sum_{\alpha \beta} \sum_{\sigma}\left(t^{\alpha \beta}_{ij} c^{\dag}_{i \alpha \sigma} c_{i \beta \sigma} + \text{h.c.} \right) + \sum_{i} \sum_{\alpha} \sum_{\sigma} \left( E_{\alpha} - \mu \right) c^{\dag}_{i \alpha \sigma} c_{i \alpha \sigma} \right] P_{s} \label{Eq:Inly_Ht}\\
H_{112-J} = & P_{s} \left[ J_{ij} \sum_{\braket{ij}} \left( \bm{S}_{i3} \cdot \bm{S}_{j3} - \frac{n_{i3} n_{j3}}{4} \right) \right] P_{s}. \label{Eq:Inly_J}
\end{align}

\enni The $i,j$ indices cover all of the sites of a \emph{three-dimensional} tetragonal lattice. $\alpha, \beta \in \{1,2,3\}$ represent the Nd $d_{z^{2}}$, $d_{xy}$, and Ni $d_{x^{2}-y^{2}}$ orbitals, respectively. We consider the tight-binding coefficients $t^{\alpha \beta}_{ij}$ and on-site energies $E_{\alpha}$ of Ref.~\onlinecite{Thomale_PRB2020}. 

The NN exchange interactions, determined by $J$, are effective only for the Ni $d_{x^{2}-y^{2}}$ orbital. Similarly, $P_{s}$ is a projection operator which eliminates doubly-occupied configurations exclusively in the Ni Hilbert space. The inter- and intra-layer NN exchanges are fixed at $J_{\parallel}=25$  meV and $J_{z}=10$ meV, respectively, as extracted by our DFT calculations. We note that the exchange interactions in the 112 compounds are smaller than those in the 438 family roughly by a factor of three. 
\subsection{Mean-field solution at $T=0$ in the slave-boson formulation}

We first discuss the case of hole-doped Ni $d_{x^{2}-y^{2}}$. The electron-doped case is discussed at the end of this section below.

In order to take into account the exclusion of doubly-occupied Ni $d_{x^{2}-y^{2}}$ states, we introduce a slave-boson representation~\cite{Kotliar}. Furthermore, we impose three conditions via (i) the standard constraint relating boson and fermion operators in the Ni $d_{x^{2}-y^{2}}$ sector due to the Gutzwiller projection, (ii) a fixed Ni $d_{x^{2}-y^{2}}$ filling, and (iii) a fixed total filling. Specifically, these conditions can be expressed as    

\enni \begin{align}
\sum_{\sigma}f^{\dag}_{i 3 \sigma}f_{i 3 \sigma} + b^{\dag}_{i} b_{i} = & 1 \label{Eq:Inly_Cndt_1}\\
\sum_{\sigma}f^{\dag}_{i 3 \sigma}f_{i 3 \sigma} = & 1-\delta \label{Eq:Inly_Cndt_2} \\
\sum_{\sigma}f^{\dag}_{i 3 \sigma}f_{i 3 \sigma} + \sum_{\alpha \neq 3} \sum_{\sigma} c^{\dag}_{i \alpha \sigma}c_{i \alpha \sigma} = & n_{Tot} \label{Eq:Inly_Cndt_3},   
\end{align}

\enni where 

\enni \begin{align}
\delta = 1-n_{3}
\end{align}

\enni is the Ni $d_{x^{2}-y^{2}}$ doping and 

\enni \begin{align}
n_{Tot} = \frac{1}{N_{s}} \sum_{i}\sum_{\alpha} \sum_{\sigma} \braket{c^{\dag}_{i\alpha} c_{i\alpha}} 
\end{align}

\enni is the total filling for the three orbitals. In order to impose these conditions we introduce associated Lagrange multipliers $\lambda_{i}, \mu_{3}$, and $\mu$. 

The $t-J$ model in the slave-boson representation is 

\enni \begin{align}
H_{112} = &  - \sum_{i<j} \sum_{\alpha,\beta \neq 3} \sum_{\sigma} 
\left( t^{\alpha \beta}_{ij} c^{\dag}_{i \alpha \sigma} c_{j \beta \sigma} + h.c. \right)  
-  \sum_{i<j} \sum_{\sigma} \left( t^{33}_{ij} f^{\dag}_{i 3 \sigma} b_{i} b^{\dag}_{j} f_{j 3 \sigma} + h.c. \right) 
\notag \\
- & \sum_{\alpha} \sum_{\sigma} \left( t^{\alpha 3}_{ij} c^{\dag}_{i \alpha \sigma} b^{\dag}_{i} f_{j \alpha \sigma} + h.c. \right) 
+  \sum_{i} \sum_{\sigma}\left( \sum_{\alpha \neq 3} E_{\alpha} c^{\dag}_{i \sigma} c_{i \sigma} + E_{3} f^{\dag}_{i \sigma} f_{i \sigma}  \right) 
\notag \\
-  & \frac{J_{ij} }{4} \sum_{\braket{ij}} \sum_{\alpha} \bigg[  \left( f^{\dag}_{i 3 \up}f^{\dag}_{j  \dn} - f^{\dag}_{i 3 \dn}f^{\dag}_{j 3 \up} \right)
\left( f_{i 3 \dn}f_{j 3 \up} - f_{i 3\up}f_{j 3 \dn} \right) 
+ \sum_{\sigma \sigma'}  f^{\dag}_{i 3\sigma}  f_{j 3 \sigma}  f^{\dag}_{j 3 \sigma'} f_{i 3 \sigma'} 
\notag \\
 - &  \sum_{\sigma} \frac{J_{ij}}{4} f^{\dag}_{i 3 \sigma} f_{i 3 \sigma} 
+  \left(1-b^{\dag}_{i} b_{i} - b^{\dag}_{j} b_{j } - b^{\dag}_{i } b_{i} b^{\dag}_{j} b_{j} \right) \bigg] 
+  \sum_{i} \lambda_{i}\left( b^{\dag}_{i}b_{i} + \sum_{\sigma} f^{\dag}_{i 3  \sigma} f_{i 3 \sigma} -1\right) \notag \\
- & \mu_{3} \sum_{i} \left( \sum_{\sigma} f^{\dag}_{i 3 \sigma} f_{i 3 \sigma} -1 + \delta \right) - \mu \sum_{i} \left[ \sum_{\sigma} \left(  \sum_{\alpha \neq 2} c^{\dag}_{i \alpha \sigma} c_{i \alpha \sigma} + f^{\dag}_{i 3 \sigma} f_{i 3 \sigma} \right) - n_{Tot} \right].
\end{align}

\enni At $T=0$, we consider solutions where the boson condenses into an uniform state such that we can replace $b^{\dag}_{i}, b_{i}$ by the real number $\braket{b_{i}}=b$ and ignore fluctuations about this state~\cite{Kotliar}. In contrast to the conventional $t-J$ model, $H$ does not conserve total boson number due to the presence of the hybridization $t^{\alpha 3}_{ij}$ between the Ni $d_{x^{2}-y^{2}}$ and Nd orbitals. Consequently, in the most general case, the boson amplitude $b$ as well as all of the orbital fillings $n_{\alpha \in \{1,2,3\}}$ must be determined self-consistently for fixed $n_{Tot}$, as for mixed-valent systems~\cite{Coleman_1987, Millis_Lee, Auerbach_Levin_PRL_1986}. Instead, we fix the Ni $d_{x^{2}-y^{2}}$ $n_{3}$ at it's value in the non-interacting case. Together with condition (i) (Eq.~\ref{Eq:Inly_Cndt_1}), this uniquely determines 

\enni \begin{align}
\braket{b^{\dag}_{i}} = & \braket{b_{i}} = b \notag \\
b^{2} = & \delta.
\end{align} 

\enni and the corresponding uniform $\lambda_{i}=\lambda$. 

We proceed to a Hartree-Fock decoupling of the exchange interactions in both the p-p and p-h channels~\cite{Brinckmann_Lee}. The corresponding Landau-Ginzburg free-energy per unit cell is

\enni \begin{align}
f = & -\frac{2T}{N_{s}} \sum_{m} \sum_{\bm{k}} \ln\left( \cosh(\beta E_{\bm{k}m})  \right) + \frac{3J_{\parallel}}{8} \left( \left|B_{x} \right|^{2} + \left|B_{y} \right|^{2} \right) + \frac{3J_{z}}{8} \left|B_{z} \right|^{2} 
+ \frac{3J_{\parallel}}{8}\left(\left|\chi_{x} \right|^{2} + \left| \chi_{y} \right|^{2} \right) + \frac{3J_{z}}{8} \left|\chi_{z} \right|^{2}
- \mu_{3} \delta.
\end{align}  

\enni In writing the free-energy per unit cell we ignored all terms corresponding to a trivial parameter-independent shift. In addition, we have absorbed the redundant Lagrange multiplier $\lambda$ into a renormalized $\mu_{3}$.  

We minimize $f$ with respect to the mean-field order parameters $B_{x,y,z}, K_{x,y,z}, \mu_{3}$, and $\mu$. While the last two are determined by fixing the $d_{x^{2}-y^{2}}$ and total fillings, respectively, the pairing and $f$-electron hopping mean-field parameters are defined as

\enni \begin{align}
B_{\bm{e}} = & \braket{f_{\bm{r}_{i}+\bm{e} 3 \dn} f_{\bm{r}_{i} 3 \up} - f_{\bm{r}_{i} + \bm{e} 3 \up} f_{\bm{r}_{i} 3 \dn}  } \notag \\
= & \frac{2}{N_{s}} \sum_{\bm{k}} \cos(\bm{k} \cdot \bm{e}) \braket{f_{\bm{k} 3 \dn}f_{\bm{-k} 3 \up}}, 
\end{align}

\enni and 

\enni \begin{align}
\chi_{\bm{e}} = & \sum_{\sigma} \braket{f^{\dag}_{\bm{r}_{i} 3 \sigma}f_{\bm{r}_{i} + \bm{e} 3 \sigma}}  
\notag \\
= & \frac{1}{N_{s}} \sum_{\bm{k}} \sum_{\sigma} e^{i\bm{k} \cdot \bm{e}} \braket{f^{\dag}_{\bm{k} 3 \sigma}f_{\bm{k} 3 \sigma}}, \label{Eq:Inly_K}
\end{align}

\enni where $\bm{e} \in \{ \bm{\hat{x}}, \bm{\hat{y}}, \bm{\hat{z}} \}$. 

$E_{\bm{k}m}$ are the eigenvalues of the effective Hamiltonian

\enni \begin{align}
H_{\bm{k}} = \begin{pmatrix}
h_{\bm{k};\alpha \beta}  & \Delta_{\bm{k}; \alpha \beta} \\
\Delta^{\dag}_{\bm{k}; \alpha \beta}  & - h^{T}_{-\bm{k};\alpha \beta} 
\end{pmatrix} 
\end{align}

\enni in a Nambu basis with spinor 
$\Psi^{T}=(c_{\bm{k} \alpha \neq 3 \up}, f_{\bm{k} 3 \up}, c^{\dag}_{-\bm{k} \alpha \neq 3 \dn}, f_{-\bm{k} 3 \dn})$. The normal part is determined by 

\enni \begin{align}
h_{\bm{k}; \alpha \beta} = \text{F.T.}\left[H_{t; \alpha \beta} (b) - \frac{3}{4} \sum_{\bm{e}} \sum_{\sigma} J_{\bm{e}} \text{Re} \left( \chi_{\bm{e}} e^{i \bm{k} \cdot \bm{e}} \right) \delta_{\alpha, 3} - \mu_{3} \delta_{\alpha, 3} \right]  ,
\end{align}

\enni where the Kronecker $\delta_{\alpha,3 }$ is not to be confused with the doping $\delta$.$H_{t}$ is given in Eq.~\ref{Eq:Inly_Ht} with tight-binding coefficients involving the Ni $d_{x^{2}-y^{2}}$ orbital re-scaled by $\sqrt{b}$ and  
where $J_{\bm{e}} = J_{\parallel}$ for $\bm{e} \in \{\bm{x}, \bm{y} \}$ and $J_{z}$ for $\bm{e}=\bm{z}$.

The pairing part is determined by 

\enni \begin{align}
\Delta_{\bm{k},\alpha \beta} = & - \sum_{\bm{e}} \frac{3J_{\bm{e}}}{4} \cos(\bm{k} \cdot \bm{e}) B_{\bm{e}} \delta_{\alpha 3} 
\notag \\
= & - \Delta_{s_{x^{2}+y^{2}}} \left[\cos(k_{x}a) + \cos(k_{y}a) \right] - \Delta_{d_{x^{2}-y^{2}}}\left[\cos(k_{x}a) - \cos(k_{y}a) \right] - \Delta_{z^{2}} \cos(k_{z}a),  
\end{align}

\enni where $a$ is the NN distance and the pairing channels which transform according to $A_{1g}$ and $B_{1g}$ representations of the point group are

\enni \begin{align}
\Delta_{s_{x^{2}+y^{2}}} = & \frac{3J_{\parallel}}{4} \left( B_{x} + B_{y}\right) \\
\Delta_{z^{2}} = & \frac{3J_{z}}{4} B_{z} 
\\
\Delta_{d_{x^{2}-y^{2}}} = & \frac{3J_{\parallel}}{4} \left( B_{x} - B_{y}\right). 
\end{align}

For the electron doped Ni $d_{x^{2}-y^{2}}$ cases, we apply the p-h transformation in Eq.~\ref{Eq:p_h}, where the constrained $d_{x^{2}-y^{2}}$ annihilation operator is replaced by 

\enni \begin{align}
P_{s} c_{i 3 \sigma} P_{s}\rightarrow \tilde{P}_{s} \tilde{c}^{\dag}_{i 3 \sigma} \tilde{P}_{s} \label{Eq:p_h}. 
\end{align} 

\enni $\tilde{P}_{s}$ projects doubly-occupied hole states in $H_{112}$. For consistency, we apply the same ph transformation to the remaining Nd operators: $c_{i \alpha \neq 3 \sigma} \rightarrow \tilde{c}^{\dag}_{i \alpha \neq 3 \sigma}$. We neglect trivial overall shifts in ground state energy and incorporate a shift in the $d_{x^{2}-y^{2}}$ on-site energy into a renormalized $\mu_{3}$. The resulting effective model is the same as in the hole-doped case provided all tight-binding coefficients change sign. We solve this model as in the hole-doped case with the constraints

\enni \begin{align}
\sum_{\sigma}\tilde{f}^{\dag}_{i 3 \sigma} \tilde{f}_{i 3 \sigma} = & 1-\tilde{\delta}  \\
\sum_{\sigma} \tilde{f}^{\dag}_{i 3 \sigma} \tilde{f}_{i 3 \sigma} + \sum_{\alpha \neq 3} \sum_{\sigma} \tilde{c}^{\dag}_{i \alpha \sigma} \tilde{c}_{i \alpha \sigma} = & \tilde{n}_{Tot} ,   
\end{align}

\enni where $\tilde{\delta}$ is the electron doping and $\tilde{n}_{Tot} = 6 - n_{Tot}$. 

The self-consistent calculations were done on a $100 \times 100 \times 100$ grid in the first 3D Brillouin Zone.

\subsection{Estimate of $T_{c}$}

In estimating the critical temperature $T_{c}$ which marks the onset of superconductivity, we follow the well-known analogous procedure in the cuprate high-$T_
{c}$ superconductors. 
 As in the case of the high-$T_{c}$ cuprates, we distinguish under- and over-doped regimes based on the Ni $d_{x^{2}-y^{2}}$ filling and apply different estimation procedures in these regimes accordingly~\cite{Lee_2006}.  

In the underdoped regime, we estimate an upper bound on $T_{c}$ via the boson condensation temperature. We note that estimates of $T_{c}$ based on suppression of Meissner screening, as in a single-orbital $t-J$ model~\cite{Lee_Wen, Ioffe_Millis}, are complicated by the presence of the Nd bands. 

As discussed in the previous section, the $t-J$ Hamiltonian for the infinite-layer compounds does not conserve total boson number due to the hybridization terms $t^{\alpha 3}_{ij},~\alpha \neq 3$. In the more general case, where only the total filling is fixed, the condensation temperature of the boson would be estimated as in the mixed-valence problem by including the effects of the hybridization terms. At the static saddle-point level, the latter are proportional to $\sum_{\alpha \neq 3} t^{\alpha 3}_{ij}\braket{f^{\dag}_{i 3 \sigma} c_{j \alpha \sigma}} + \text{h.c.}$~\cite{Coleman_1987, Millis_Lee, Auerbach_Levin_PRL_1986}. Since the hybridization strengths $t^{\alpha 3}_{ij}$ are smaller than the NN in-plane hopping $t^{33}_{ij}$ by an order of magnitude, we ignore the contribution of the former in determining the boson condensation temperature. In our calculations, we also fix the average boson density by fixing the Ni $d_{x^{2}-y^{2}}$ filling. The bosons are essentially free with NN tight-binding coefficients determined by the $T=0$ mean-field calculation~\cite{Kotliar}. The effective boson Hamiltonian reads

\enni \begin{align}
H_{B} = & - \sum_{\bm{k}} 
\left(  \sum_{ \bm{e}} 2 t_{B \bm{e}} \cos(\bm{k} \cdot \bm{e}) - \lambda - J \right) b^{\dag}_{\bm{k}} b_{\bm{k}},  
\end{align}
 
\enni together with the constraint 

\enni \begin{align}
\braket{b^{\dag}_{i} b_{i}} = & \delta 
\notag \\
\approx & \braket{b^{\dag}_{i} b_{j}}.
\end{align}

\enni The boson tight-binding coefficient is 
 
\enni \begin{align}
t_{B \bm{e}} = t^{33}_{\bm{e}} \chi_{\bm{e}}  + \frac{J_{\bm{e}}}{4} \delta,
\end{align}

\enni where $\bm{e} \in \{\bm{\hat{x}} , \bm{\hat{y}}, \bm{\hat{z}}\}$ and we used the fact that the $\chi_{\bm{e}}$ are real in the mean-field solution. Near the Ni $d_{x^{2}-y^{2}}$ half-filling point, we take $K_{\bm{e}}$ (Eq.~\ref{Eq:Inly_K}) to be equal to its $T=0$ value since we anticipate that $T_{RVB}$ associated with $\chi$ is finite as $\delta \rightarrow 0$~\cite{Kotliar}. At the transition, the boson chemical potential vanishes as 

\enni \begin{align}
\lambda \rightarrow 2 \sum_{\bm{e}} t_{B \bm{e}} - J.
\end{align}

\enni We determine $T_{c}=1/\beta_{c}$ by imposing the boson density constraint

\enni \begin{align}
\delta = & \frac{1}{N_{s}} \sum_{\bm{k}} n_{B \bm{k}},
\end{align}

\enni where $n_{B \bm{k}}$ is the Bose-Einstein factor 

\enni \begin{align}
n_{B \bm{k}} = & \frac{1}{e^{\beta_{c} \epsilon_{B \bm{k}}}-1}.
\end{align}

In the over-doped regime, we expect that $T_{c}$ can be estimated from BCS theory as~\cite{Won_Maki}

\enni \begin{align}
T_{c} \approx  \left| \Delta_{d} \right| /2.14,
\end{align}

\enni where $\left| \Delta_{d} \right|$ is the amplitude of the d-wave pairing at $T=0$.

\subsection{Hartree mean-field parameters at $T=0$}

\begin{figure}[ht!]
\resizebox{0.5\columnwidth}{!}{\input{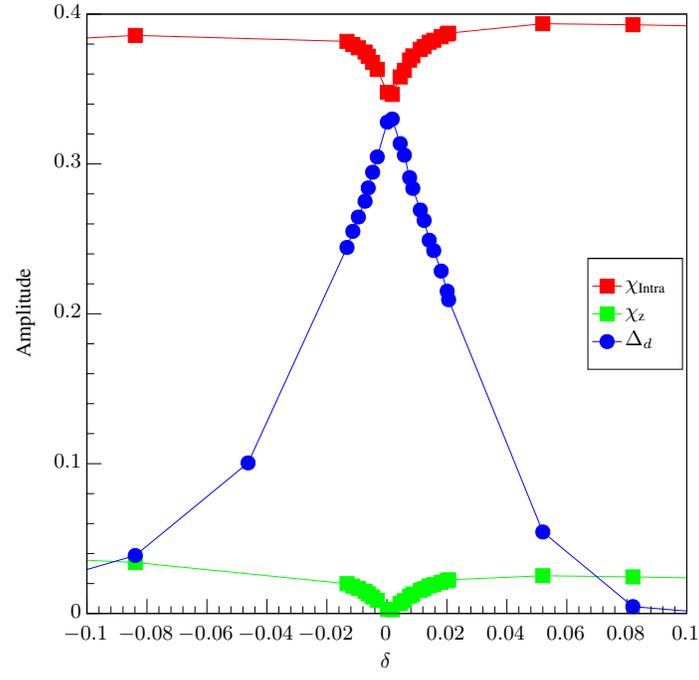}}
\caption{Dimensionless intra-layer Hartree mean-field parameter, the Hartree mean-field parameter along z, and dimensionless $d_{x^{2}-y^{2}}$ pairing amplitude for the Ni orbital at $T=0$ as functions of doping for the same orbital in 112 compounds.}
\label{Fig:2_SM}
\end{figure}

As for the case of the 438 compounds, we find that all of the Hartree mean-field parameters are real and that they obey 

\enni \begin{align}
\chi_{\bm{\hat{x}}} = & \chi_{\bm{\hat{y}}} 
= \chi_{\text{Intra}}.
\end{align}

We plot these along with $\chi_{z}$ and the $d_{x^{2}-y^{2}}$ dimensionless pairing 

\enni \begin{align}
\Delta_{d} = & \left|B_{x} -B_{y} \right|,
\end{align} 

\enni in Fig.~\ref{Fig:2_SM}. As in the 438 case and the single-band $t-J$ model~\cite{Kotliar}, we find that $\chi_{\text{Inter}} \approx \Delta_{d}$ near half-filling. In addition, $\chi_{z}$ is strongly suppressed in this regime. 


\end{document}